# Mixing of $0^+$ and $0^-$ observed in hyperfine and Zeeman structure of ultracold $Rb_2$ molecules

**Markus Deiß, Björn Drews, and Johannes Hecker Denschlag**

Institut für Quantenmaterie and Center for Integrated Quantum Science and Technology IQ$^{ST}$, Universität Ulm, 89069 Ulm, Germany

**Eberhard Tiemann**

Institut für Quantenoptik, Leibniz Universität Hannover, 30167 Hannover, Germany

E-mail: `johannes.denschlag@uni-ulm.de`

**Abstract.** We study the combination of hyperfine and Zeeman structure in the spin-orbit coupled $A^1\Sigma_u^+ - b^3\Pi_u$ complex of $^{87}Rb_2$. For this purpose, absorption spectroscopy at a magnetic field around $B = 1000\,G$ is carried out. We drive optical dipole transitions from the lowest rotational state of an ultracold Feshbach molecule to various vibrational levels with $0^+$ symmetry of the $A-b$ complex. In contrast to previous measurements with rotationally excited alkali-dimers, we do not observe equal spacings of the hyperfine levels. In addition, the spectra vary substantially for different vibrational quantum numbers, and exhibit large splittings of up to 160 MHz, unexpected for $0^+$ states. The level structure is explained to be a result of the repulsion between the states $0^+$ and $0^-$ of $b^3\Pi_u$, coupled via hyperfine and Zeeman interactions. In general, $0^-$ and $0^+$ have a spin-orbit induced energy spacing $\Delta$, that is different for the individual vibrational states. From each measured spectrum we are able to extract $\Delta$, which otherwise is not easily accessible in conventional spectroscopy schemes. We obtain values of $\Delta$ in the range of $\pm 100$ GHz which can be described by coupled channel calculations if a spin-orbit coupling is introduced that is different for $0^-$ and $0^+$ of $b^3\Pi_u$.

## 1. Introduction

The strongly spin-orbit coupled $A^1\Sigma_u^+ - b^3\Pi_u$ complex of alkali-metal dimers has been studied in great detail in recent years, stimulated by the fruitful combination of high-resolution spectroscopy and numerical close-coupled calculations. Various homonuclear ($Rb_2$ [1, 2, 3], $Cs_2$ [4, 5], $Na_2$ [6, 7], $K_2$ [8, 9, 10, 11], $Li_2$ [12, 13]) and heteronuclear (NaRb [14, 15], RbCs [16, 17, 18], KRb [19], NaCs [20], KCs [21, 22, 23], NaK [24, 25, 26]) species have been investigated and modeled. Potential energy curves as well as $r$-dependent spin-orbit-coupling functions were extracted, where $r$ is the internuclear separation. Concerning the hyperfine structure of the $A-b$ state, however, only little experimental data is available so far.

For thermal and thus rotationally excited samples of $Na_2$ and $K_2$ hyperfine structures with line splittings up to hundreds of MHz, characterized by nearly equidistant separations of the energy levels were observed [7, 11]. Such hyperfine structures of the $\Omega = 0$ components of the $A-b$ complex come about owing to the molecular rotation that mixes different $\Omega$ components. For the case of low rotational angular momentum $J$, line splittings of at most a few MHz are expected. Indications of such small hyperfine splittings for $J = 1$ RbCs molecules in state $\Omega = 0$ were reported in Ref. [16], but a detailed analysis was not given.

In this work, we investigate the combined hyperfine and Zeeman pattern of the $A-b$ complex for $Rb_2$ molecules with $J = 1$ observed by exciting an appropriate Feshbach molecular state [see level scheme in Fig. 1(a)]. Particularly for states, where the main component exhibits $b^3\Pi_u\ 0^+$ symmetry, we measure large level spacings of up to 160 MHz. Furthermore, the line pattern is not equally spaced and the overall structure changes strongly from one vibrational level to another. Consequently, our spectra are dominated by a mechanism different from the one discussed previously in the context of fast rotating molecules. In fact, we find that the observed energy level structures corresponding to vibrational states of $b^3\Pi_u\ 0^+$ arise from second order hyperfine and Zeeman interaction coupling the $0^+$ and $0^-$ components of $b^3\Pi_u$. More precisely, these two interactions work together in a cooperative way enhancing the effect. By fitting a relatively simple model to our data we extract the initially unknown frequency spacing $\Delta$ between $0^-$ and $0^+$ for each vibrational level. This is an important result of our work because the state $b^3\Pi_u\ 0^-$ is not directly accessible in spectroscopy schemes starting from any singlet or triplet ground state molecular level. Our derived values for $\Delta$ systematically deviate by about 90 GHz from predictions of close-coupled channel calculations. We interpret this as a difference in the spin-orbit coupling function for $0^-$ and $0^+$.

This article is organized as follows. In section II, we give an overview of the experimental setup and the spectroscopy scheme. Then, section III describes the relevant molecular energy states needed for the presentation of our experimental results in section IV. In section V we introduce a simple model that fully explains the characteristics of the observed spectra. Our model calculations are discussed in section VI along with the interpretation of the data and the determination of $\Delta$ for the investigated vibrational states of $b^3\Pi_u$. Finally, in section VII we describe the extension of the potential scheme needed for modeling the observations by coupled channel calculations.

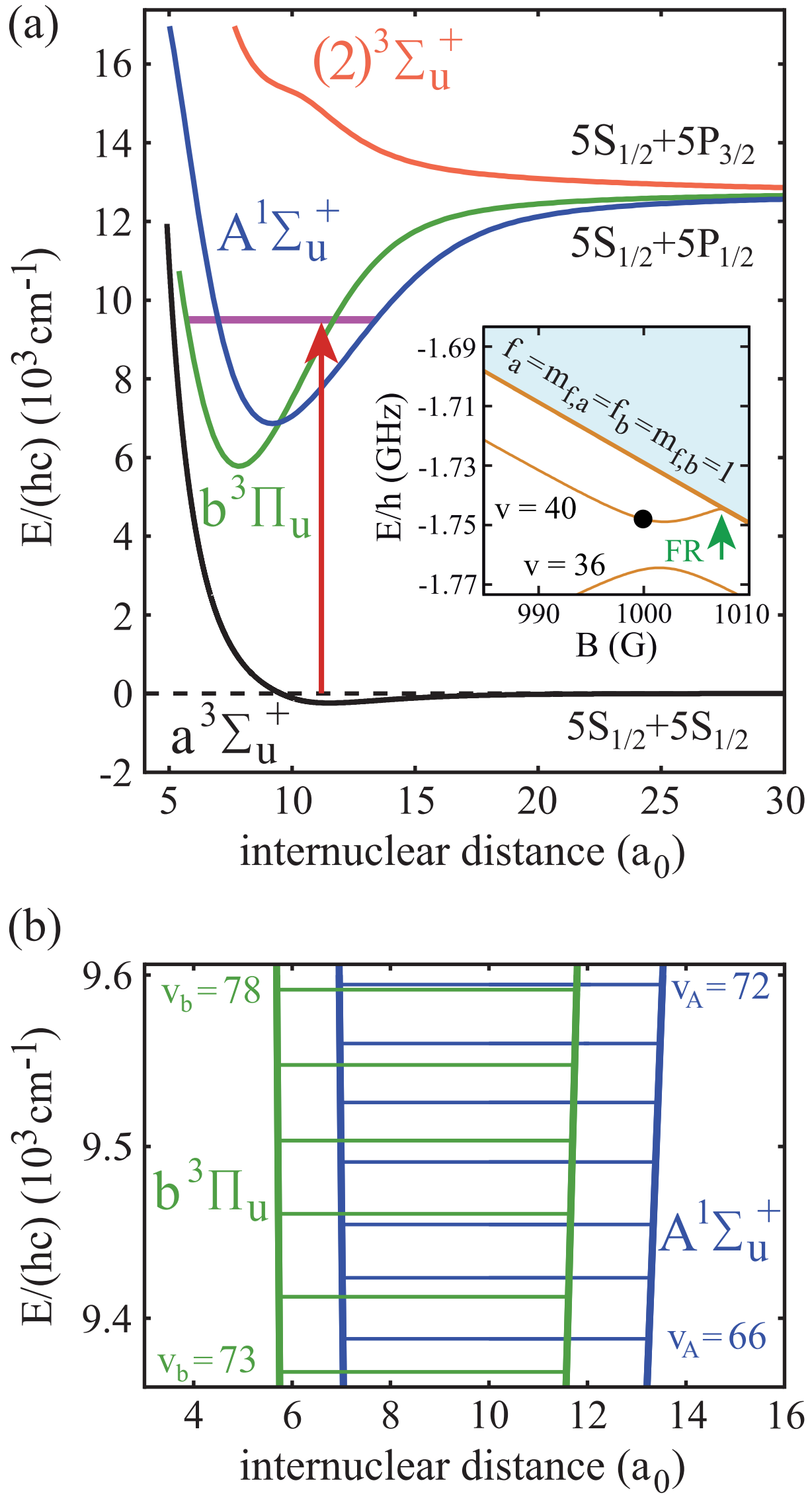

**Figure 1.** (a) Spectroscopy scheme. Weakly bound Feshbach molecules are irradiated by a laser pulse and excited to molecular levels of the $A^1\Sigma_u^+ - b^3\Pi_u$ manifold from where they spontaneously decay to nonobserved states. The potential $(2)^3\Sigma_u^+$ is included because it couples to $0^-$ of $b^3\Pi_u$ (see text). Furthermore, the inset shows the level structure in the vicinity of the Feshbach resonance (FR). At a magnetic field of $B = 999.9\,$G the Feshbach state (indicated by the black circle) is located $1.748\,\text{GHz} \times h$ below the $|f_a = 1, m_{f_a} = 1\rangle + |f_b = 1, m_{f_b} = 1\rangle$ dissociation threshold at 0 G. In (b), the vibrational levels $\text{v}_A$ and $\text{v}_b$ within the $A^1\Sigma_u^+ - b^3\Pi_u$ $0^+$ complex that are relevant for our measurements are depicted. All potential curves are taken from [27], while the energies of $\text{v}_A$ and $\text{v}_b$ correspond to the calculated values given in [28] (see also tables 1 and 2).

## 2. Experimental setup

We carry out our experiments with a pure sample of about $N_0 = 3\times10^4$ weakly bound $^{87}Rb_2$ Feshbach molecules which have both $X^1\Sigma_g^+$ and $a^3\Sigma_u^+$ character. The setup and the molecule preparation scheme are described in detail in Refs. [29, 30]. Therefore, they are just briefly presented here. Initially a BEC or ultracold thermal cloud of spin-polarized $^{87}$Rb atoms with total angular momentum $f = 1$, $m_f = 1$ is loaded into a rectangular, 3D optical lattice at a wavelength of $\lambda = 1064.5$ nm, which is formed by a superposition of three linearly polarized standing light waves with polarizations orthogonal to each other. By slowly crossing the magnetic Feshbach resonance (FR) at 1007.4 G from high to low fields, pairs of atoms in doubly occupied lattice sites are converted into weakly bound molecules. Afterwards, the magnetic field is set to $999.9\pm0.1$ G, where we perform the spectroscopy. In order to get rid of remaining atoms, a combined microwave and light pulse is applied which removes them from the lattice. We end up with a pure ensemble of molecules that resides in the lowest Bloch band of the optical lattice with no more than a single dimer per lattice site. The lattice depth for the molecules with respect to each of the standing light waves of the optical lattice is about $64E_R$ where $E_R = h^2/(2m\lambda^2)$ represents the recoil energy. Here, $m$ denotes the mass of the molecule and $h$ is Planck's constant. Since at these lattice depths the tunneling rate is very small, intermolecular collisions are strongly suppressed, and we measure lifetimes on the order of 1 s.

Figure 1(a) shows the spectroscopy scheme. The Feshbach molecule ensemble is irradiated by a rectangular light pulse for a duration $\tau$ of typically a few ms. At the location of the molecular sample the beam waist is about 1.1 mm. For the observed spectra, we used laser powers of tens or hundreds of $\mu$W. The light propagates orthogonally to the quantization axis which is defined by the applied magnetic field that points in vertical direction. By using a half-wave plate we can choose the light being polarized either in the horizontal plane or in the vertical axis giving rise to $\sigma$ transitions (i.e., $\sigma^+$ and $\sigma^-$) or $\pi$ transitions. Molecules, that are resonantly excited from the Feshbach state to a level of the $A-b$ complex, are in general lost due to subsequent fast decay to nonobserved states. We measure the remaining fraction $N/N_0$ of Feshbach dimers. For this purpose, we dissociate the molecules by ramping back over the Feshbach resonance and detect the corresponding atom number via absorption imaging.

The spectroscopy is performed at wavelengths between 1042 and 1068 nm (corresponding to about $9360-9600\,\mathrm{cm}^{-1}$) using a grating-stabilized cw diode laser that has a short-term linewidth of $\sim100$ kHz. This laser is frequency-stabilized to a Fizeau interferometer wavemeter (High Finesse WS7), with an update rate of about 10 Hz. As the laser frequency drifts between updates, we obtain a frequency stability of $\pm(2-5)$MHz. The wavemeter is calibrated to an atomic $^{87}$Rb reference signal at 780 nm in intervals of minutes. It has a specified absolute accuracy of 60 MHz, but the accuracy is on the MHz level for difference frequency determinations within several hundred MHz. Furthermore, over a period of several months we checked the frequency readings of the wavemeter for the same molecular transitions and did not find deviations of more than $\pm10$ MHz. This demonstrates the good reproducibility of the wavemeter readings in connection with the calibration mentioned above.

## 3. Relevant states

### *3.1. Feshbach molecules*

The $Rb_2$ Feshbach molecules in our experiment are weakly bound dimers with both singlet and triplet character, i.e., the selected state is a mixture of $X^1\Sigma_g^+$ and $a^3\Sigma_u^+$ ([30, 31]). However, only the $X^1\Sigma_g^+$ component allows to drive transitions to the $A-b$ complex because for an electric dipole transition the $u/g$ symmetry has to change and the $A-b$ complex has $u$ symmetry. According to coupled channel calculations, at a magnetic field of 999.9 G the singlet component, mainly characterized by $S=L=R=0$, $I=2$, $m_I=2$, $F=2$, contributes 16% to the Feshbach state which has the exact quantum numbers $m_F=2$ and parity $+$. Here, $S$, $L$, $R$, $I$ and $F$ ($\vec{F}=\vec{R}+\vec{L}+\vec{S}+\vec{I}$) denote the quantum numbers of the total electronic spin, the total orbital angular momentum, the rotation of the atom pair, the total nuclear spin, and the total molecular angular momentum, respectively. Furthermore, $m_I$ and $m_F$ represent the corresponding projections onto the quantization axis. Consequently, the singlet component of the Feshbach molecules has $J=0$ ($\vec{J}=\vec{R}+\vec{L}+\vec{S}$).

The inset of Fig. 1(a) shows the molecular level structure in the vicinity of the Feshbach resonance. Throughout the present work, all excitation energies are given with respect to the $|f_a=1, m_{f_a}=1\rangle+|f_b=1, m_{f_b}=1\rangle$ atomic dissociation limit at 0 G. Note, its energy is $8.543\,\mathrm{GHz}\times h$ below the atomic dissociation limit when hyperfine interaction is ignored. At a magnetic field of 999.9 G the Feshbach state is located at $-1.748\,\mathrm{GHz}\times h$. Here, the main contribution is determined by the Zeeman shift of the atom pair $|f_a=1, m_{f_a}=1\rangle+|f_b=1, m_{f_b}=1\rangle$. The molecular binding energy is only about $20\,\mathrm{MHz}\times h$ with respect to this threshold.

### *3.2. $A^1\Sigma_u^+ - b^3\Pi_u$ complex*

Spin-orbit interaction leads to a mixing of the states $A^1\Sigma_u^+$ and $b^3\Pi_u$ forming the $A-b$ complex. In a simple approach this mixing comes about in two steps. First, due to spin-orbit coupling the state $b^3\Pi_u$ splits up into three components, $\Omega=0,1,2$. The quantum number $\Omega$ denotes the projection of the sum of all electronic angular momenta onto the internuclear axis and equals the projection of the molecular angular momentum $J$ on the same axis. For $Rb_2$ the relative separation of the three terms is about $80\,\mathrm{cm}^{-1}$, as mainly determined by the atomic spin-orbit splitting of Rb in its $5^2P$ state. At this stage, the $b^3\Pi_u$, $\Omega=0$ state has two degenerate components, $0^+$ and $0^-$. Second, spin-orbit coupling mixes $A^1\Sigma_u^+$ (i.e., $0^+$ symmetry) and $b^3\Pi_u$ $0^+$, whereas the $b^3\Pi_u$ $0^-$ component couples to $(2)^3\Sigma_u^+$ $0^-$ [see Fig. 1(a)]. As a consequence of the repulsive interactions $0^+$ and $0^-$ of $b^3\Pi_u$ are separated from each other, which is referred to as $\Lambda$-type splitting [32]. This effect is crucial for the interpretation of the observations of the present work.

The vibrational levels of the $A-b$ states relevant to our measurements are illustrated in Fig. 1(b). The levels with dominant triplet (singlet) character are indicated by vibrational quantum numbers $\mathrm{v}_b$ ($\mathrm{v}_A$). Moreover, tables 1 and 2 list the numerical values for the term energies and the $b$ state admixtures calculated by Drozdova *et al.* [1] and taken from [28]. The calculation is based on a two-potential approach considering $A^1\Sigma_u^+$ and $b^3\Pi_u$ ($\Omega = 0^+, 1, 2$). The mixing is described by the parameter $p_b$, which represents the probability of finding the vibrational level in the electronic state $b$. Consequently, for the $A$ state the corresponding parameter is given by $p_A = 1 - p_b$. All other admixtures like $\Delta\Omega = 1$ are negligible in our cases. Our spectroscopy scheme addresses only the $A$ component of a vibrational level of the $A-b$ manifold. Moreover, only states with angular momentum $J=1$ and negative parity can be observed, because the electronic singlet component of the Feshbach molecule has the quantum number $J=0$ and positive total parity.

**Table 1.** Comparison of calculated ($E_{\mathrm{calc}}$) and measured ($E_{\mathrm{exp}}$) level energies for various vibrational levels $\mathrm{v}_A$ of the $A^1\Sigma_u^+$ state with $J=1$. All level energies $E_{\mathrm{exp}}$ are observed with $\pi$-polarized light. The column $\varepsilon = (E_{\mathrm{calc}} - E_{\mathrm{exp}})/hc$ gives the difference of the measured and predicted values. Furthermore, the parameter $p_b$ denotes the admixture of the $b^3\Pi_u$ potential and $\delta$ represents the measured frequency difference between the $\sigma$ and the $\pi$ resonance. For the case of $\mathrm{v}_A = 67$, 68 and 70 we only performed spectroscopy using $\pi$-polarized light and therefore $\delta$ was not determined. The values for $p_b$ and $E_{\mathrm{calc}}$ are taken from [28].

| $\mathrm{v}_A$ | $p_b$ [%] | $E_{\mathrm{calc}}/(hc)$ [cm$^{-1}$] | $E_{\mathrm{exp}}/(hc)$ [cm$^{-1}$] | $\varepsilon$ [$10^{-3}$cm$^{-1}$] | $\delta$ [MHz] |
|---|---|---|---|---|---|
| 66 | 15.60 | 9388.005 | 9387.9967 | 8.3 | -2 |
| 67 | 28.50 | 9423.589 | 9423.5794 | 9.6 | |
| 68 | 23.21 | 9454.571 | 9454.5652 | 5.8 | |
| 69 | 7.66 | 9491.049 | 9491.0451 | 3.9 | -2 |
| 70 | 10.82 | 9525.742 | 9525.7346 | 7.4 | |
| 72 | 39.36 | 9594.454 | 9594.4485 | 5.5 | -22 |

**Table 2.** Comparison of calculated ($E_{\mathrm{calc}}$) and measured ($E_{\mathrm{exp}}$) level energies for various vibrational levels $\mathrm{v}_b$ of the $b^3\Pi_u$ $0^+$ state with $J=1$, analogous to table 1. The parameter $\Delta$ is the splitting of the $0^\pm$ components as determined by fitting our theoretical model to the measured spectra (see section 6).

| $\mathrm{v}_b$ | $p_b$ [%] | $E_{\mathrm{calc}}/(hc)$ [cm$^{-1}$] | $E_{\mathrm{exp}}/(hc)$ [cm$^{-1}$] | $\varepsilon$ [$10^{-3}$cm$^{-1}$] | $\Delta$ [GHz] |
|---|---|---|---|---|---|
| 73 | 82.70 | 9368.758 | 9368.7480 | 10.0 | $81.8^{+10.6}_{-8.5}$ |
| 74 | 69.59 | 9412.519 | 9412.5122 | 6.8 | $104.5^{+23.9}_{-16.4}$ |
| 75 | 73.80 | 9460.874 | 9460.8718 | 2.2 | $-19.7^{+0.6}_{-0.6}$ |
| 76 | 88.37 | 9503.516 | 9503.5040 | 12.0 | $40.0^{+2.2}_{-2.0}$ |
| 78 | 58.25 | 9591.479 | 9591.4721 | 6.9 | $36.4^{+3.4}_{-2.9}$ |

## 4. Experimental observations

### 4.1. Spectra of A levels

We first discuss the data for levels with mainly $A^1\Sigma_u^+$ character. Six different vibrational states ($\mathrm{v}_A = 66$ to 70 and 72) have been investigated. The obtained spectra for $\mathrm{v}_A = 66$ to 70 look very similar. Figure 2(a) shows the recording for $\mathrm{v}_A = 66$ as an example. Two resonance dips are visible, one being the $\pi$ transition ($\Delta m_F = 0$), while the other one is the $\sigma$ transition ($\Delta m_F = \pm 1$). Within the measurement uncertainty of a few MHz both resonances are located on top of each other and Zeeman or hyperfine splitting is not observed. We determine the transition frequencies from fits to the data using the function $e^{-KL}$, where the amplitude $K$ is a free fitting parameter and $L$ represents a Lorentzian. Typically, the obtained transition linewidths (FWHM) are on the order of $10-20\,\mathrm{MHz}$.

In table 1 the absolute energies of states $\mathrm{v}_A$ derived from the $\pi$ resonances are summarized and compared to theoretical predictions. The admixing parameter $p_b$ and $E_{\mathrm{calc}}$ are taken from Ref. [28]. Since, the calculations were originally given with respect to the potential minimum of $X^1\Sigma_g$, for the comparison to our experimental results, we added the electronic term energy $T_e^X = -3993.5928(30)\,\mathrm{cm}^{-1}/(hc)$ of $X^1\Sigma_g^+$ [31] and the hyperfine shift of $+8.543\,\mathrm{GHz} \times h/c$, where $c$ is the speed of light. The overall agreement between the theoretical and experimental data is within the uncertainty of the theoretical predictions of $0.01\,\mathrm{cm}^{-1}/(hc)$ (corresponding to $300\,\mathrm{MHz} \times h$). Noticeably, the calculated values are systematically higher by several $10^{-3}\,\mathrm{cm}^{-1}$ compared to our measurements. Besides a possible systematic uncertainty within the theoretical model, these deviations can also arise from the limited accuracy of our wavemeter and the uncertainty of the energy $T_e^X$.

In contrast to the states $\mathrm{v}_A = 66$ and 69, where both, the $\pi$ and the $\sigma$ transition occur at the same frequency within the measurement uncertainty, $\mathrm{v}_A = 72$ shows a significant splitting [cf. Fig. 2(b)]. This is due to the fact that the admixing of the $b$ state is relatively large ($p_b \approx 40\%$, see table 1) and a $b^3\Pi_u\ 0^-$ level is located energetically close-by. The level $\mathrm{v}_A = 72$ significantly exhibits the characteristics of $b^3\Pi_u\ 0^+$, which will be discussed in the following sections.

### 4.2. Spectra of b levels

Our spectroscopic data on states with mainly triplet character, i.e., $p_b > 50\%$, are shown in Fig. 3. For all investigated vibrational quantum numbers $\mathrm{v}_b = 73$ to 76 and 78 we only clearly observe a single resonance dip when using $\pi$-polarized light. Contrary to that, the scans related to $\sigma$ polarization reveal 2 or 3 resonance features of which some might have an unresolved substructure. In each spectrum, the $\sigma$ transitions are well separated from the $\pi$ transition. We therefore choose the $\pi$ resonance as a local reference to which we assign the frequency $\delta = 0$ in the figures.

At first sight the spectra for $\mathrm{v}_b = 73$ to 76 and 78 might look somewhat irregular. For different vibrational levels $\mathrm{v}_b$ the number of transitions, their splittings, and their relative intensities vary. In addition, the splittings for a given $\mathrm{v}_b$ are not equidistant as mentioned earlier for high $J$. However, closer inspection reveals that all these spectra are characterized by a similar pattern. To show this, we arrange the spectra in the order $\mathrm{v}_b = 74$, 73, 78 and 76 [Fig. 3(a)-(d)] corresponding to their respective splitting magnitude. In each spectrum the $\sigma$ lines are located at $\delta > 0$. There is always one weak resonance next to the $\pi$ transition and one strong resonance feature at larger $\delta$. For $\mathrm{v}_b = 76$, where the total splitting is very large, the resonance dip at $\delta \sim 90\,\mathrm{MHz}$ seems to split up into two or more lines. Due to the limited resolution of about 5 MHz in our experiment we cannot clearly resolve the individual resonance lines, but the observed fluctuations in the number of molecules are a clear indication of an internal structure of this resonance dip.

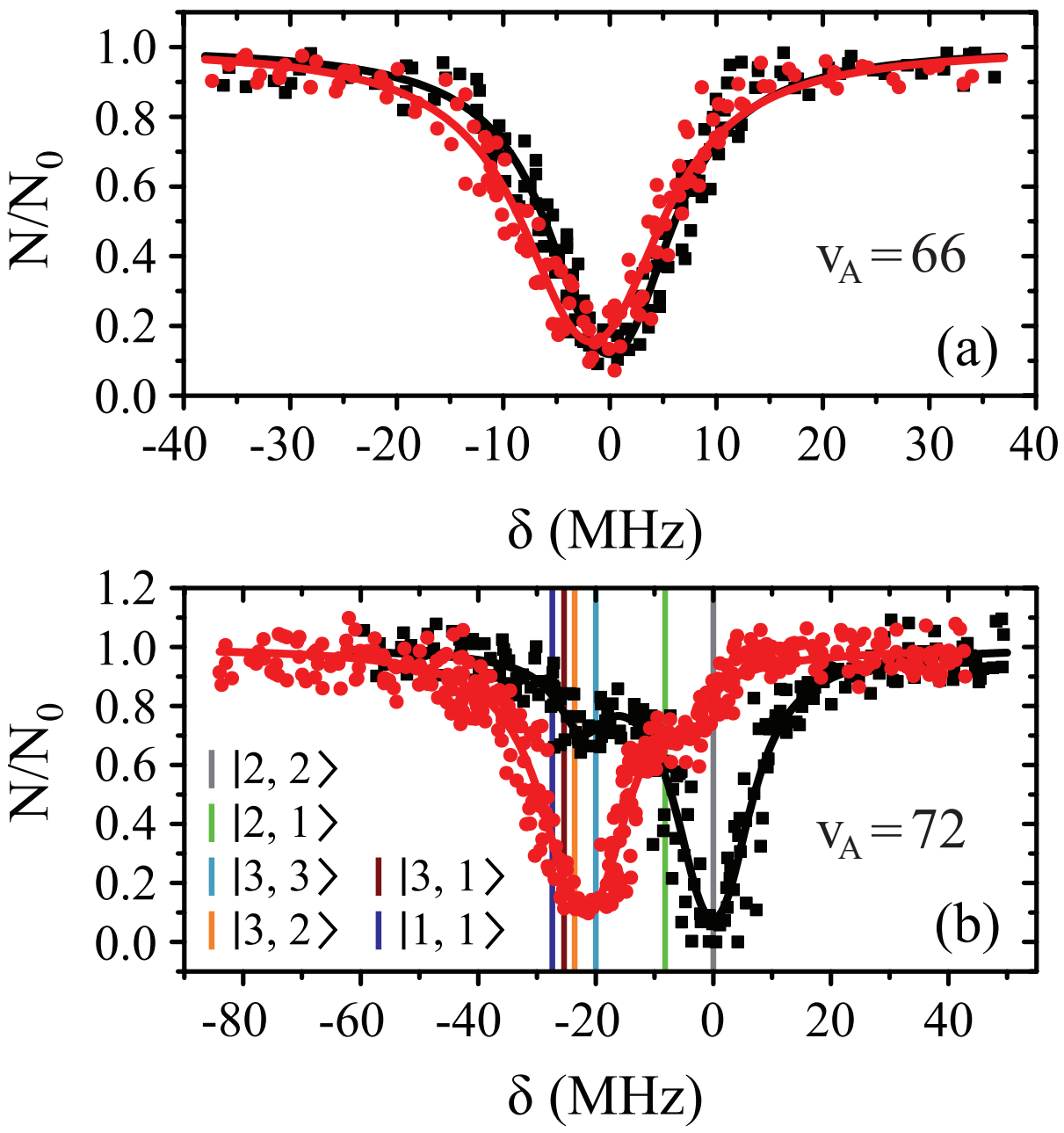

**Figure 2.** (Loss resonances for excitation of molecules from the Feshbach state to vibrational levels $\mathrm{v}_A = 66$ (a) and $\mathrm{v}_A = 72$ (b) of the $A^1\Sigma_u^+$ potential obtained with $\pi$-polarized light (black squares) and $\sigma$-polarized light (red circles). Shown is the fraction $N/N_0$ of remaining Feshbach dimers dependent on the detuning $\delta$, where $\delta = 0$ is at the resonance frequency of the strong $\pi$ transition. The corresponding offset energies are listed in table 1. Solid lines are fits of the function $e^{-KL}$ to the data (see section 4.1). For a given vibrational quantum number $\mathrm{v}_A$ the measurements with $\pi$- and $\sigma$-polarized light are performed using the same laser intensities and pulse lengths. Colored vertical lines in part (b) indicate the frequency positions of the levels $|F', m_F'\rangle$ resulting from our model calculations (see section 6).

In contrast, the spectrum of $\mathrm{v}_b = 75$ [Fig. 3(e)] is inverted compared to the spectra discussed before and exhibits three $\sigma$ resonances, all of them at $\delta < 0$. As it is spread over an even larger frequency range of about 160 MHz, the resonance dips at $-160\,\mathrm{MHz}$ and $-130\,\mathrm{MHz}$ are clearly separated from each other. The offset energies for the observed lines at $\delta = 0$ are listed in table 2 and are compared to the theoretical predictions of Ref. [28]. Again, the agreement is within the theoretical uncertainty. However, we note that the measured $\pi$ transitions contain shifts due to hyperfine and Zeeman interaction. These shifts of up to 190 MHz (see section 6) would need to be subtracted for a proper comparison of the data

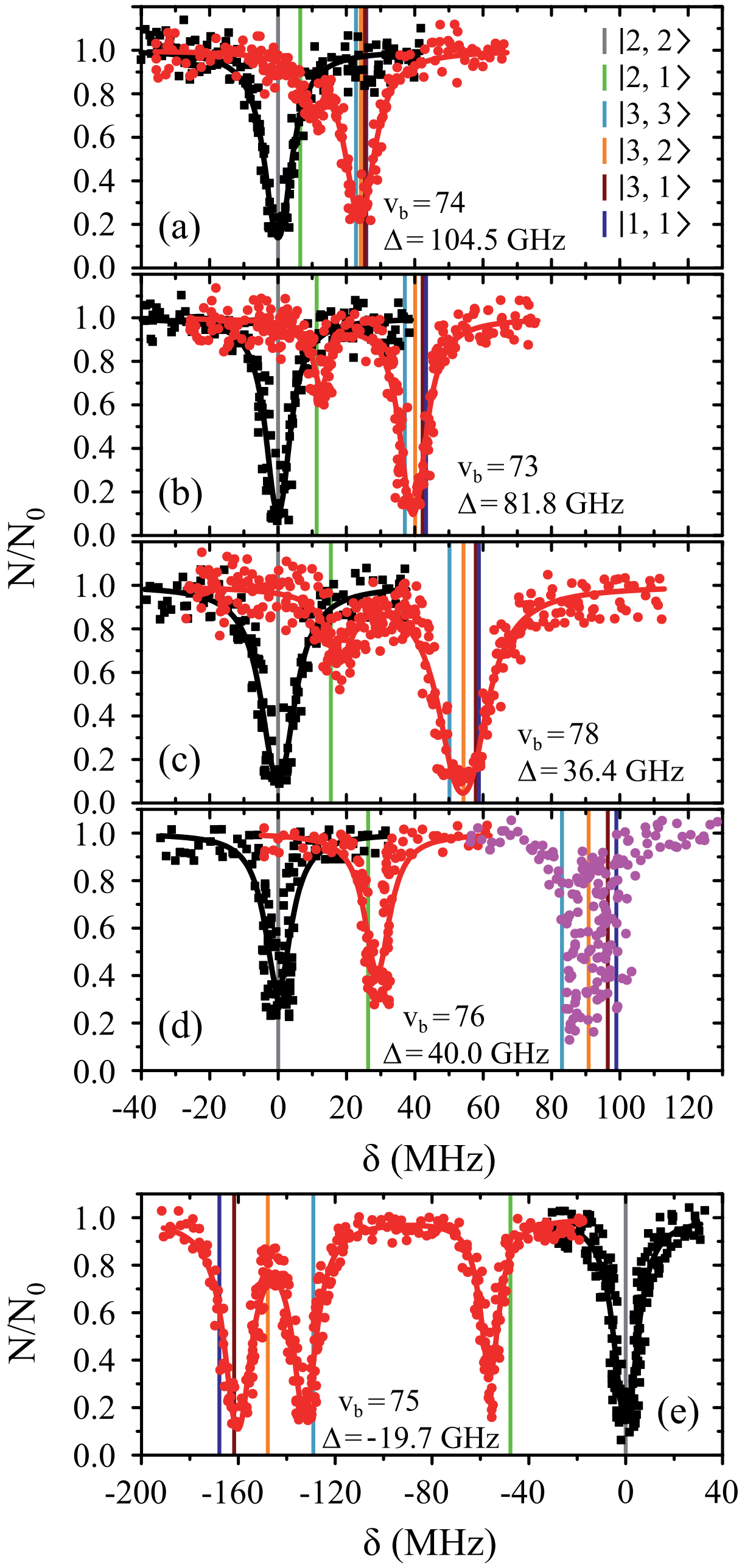

**Figure 3.** Loss spectra for excitation of molecules from the Feshbach state to vibrational levels $v_b = 74$ (a), $v_b = 73$ (b), $v_b = 78$ (c), $v_b = 76$ (d), and $v_b = 75$ (e) of the $b^3\Pi_u$ potential obtained with $\pi$-polarized light (black squares) and $\sigma$-polarized light (red circles). All parameter denotations, the fit function and the meanings of the vertical lines are identical to those of Fig. 2. The offset energies corresponding to the transitions at $\delta = 0$ are given in table 2. For $v_b = 73$ and 74 the intensity and pulse length of the $\sigma$-polarized light was the same as for $\pi$ polarization. The spectra of $v_b = 78$ ($v_b = 75$) were measured with different pulse lengths $\tau$, where the ratio was $\tau_\sigma/\tau_\pi = 5/3$ ($\tau_\sigma/\tau_\pi = 2/1$). Concerning $v_b = 76$, data of two scans with $\sigma$-polarized light are shown (magenta and red). Whereas the magenta data points were obtained using the same pulse area as for $\pi$ polarization, it was by a factor of eight larger when measuring the red data points.

with the calculations of Ref. [1].

## 5. Simple model of the molecule

In principle, hyperfine and Zeeman interaction within the $^3\Pi$ state of diatomic molecules has been theoretically investigated in depth (see, e.g., the 4th-order perturbation approach of [33]). However, properly applying such theoretical (and often complex) approaches to interpret measured spectra can still be a challenge because of the large number of parameters for representing the different orders. Therefore, we have developed a simple model which neglects some fundamental properties of a molecule. Nevertheless, it should be adequate to explain semi-quantitatively the Zeeman and hyperfine structure observed in our spectra.

In our model, the $Rb_2$ molecule is treated as a rigid rotor with fixed internuclear separation. Consequently, there is no vibrational degree of freedom. However, the positions of the nuclei can be interchanged. This is necessary in order to construct fully antisymmetric wave functions for the system of the two nuclei and the two valence electrons owing to the particles' fermionic character. Essentially, we consider the molecule as if it was composed of two unperturbed neutral atoms, of which, however, the angular momenta $\vec{L}$ and $\vec{J}$ are strongly coupled to the rigid rotator axis. In each of the atoms the orbital angular momentum $L_i$ of the local electron $i$ is a good quantum number. Thus, molecules belonging to the atom pair $5S_{1/2}+5P_{1/2}$ have both a $p$-orbital with $L_i = 1$ and a $s$-orbital with $L_i = 0$, and the total orbital angular momentum is $L = 1$ ($\vec{L} = \vec{L}_1 + \vec{L}_2$). Therefore, the two valence electrons can never be found in the same orbital. Coupling the electrons to the rotator axis (which corresponds to the internuclear axis) forms the electronic states $^{2S+1}\Lambda_{u/g}$ of the molecule. For simplicity, in the following discussion we restrict the model to those electronic states that are most relevant to describe our observations, i.e., states with $u$-symmetry and $\Omega = 0$. These are $b^3\Pi_u$ ($0^+$ as well as $0^-$), $A^1\Sigma_u^+$ and $(2)^3\Sigma_u^+$ (see section 3.2).

The molecule is described by the Hamiltonian

$$H = H_{\text{Diag}} + H_{\text{SO}} + H_{\text{R}} + H_{\text{HF}} + H_{\text{Z}}, \tag{1}$$

which, in addition to a diagonal energy matrix, contains spin-orbit coupling, nuclear rotation, hyperfine and Zeeman interaction. The diagonal energy matrix

$$H_{\text{Diag}} = \sum_{\Lambda,S,0^\pm} E_{\Lambda,S,0^\pm}\, P(^{2S+1}\Lambda_u\, 0^\pm) \tag{2}$$

sets the initial values for the energies $E_{\Lambda,S,0^\pm}$ of the electronic levels $^{2S+1}\Lambda_u\, 0^\pm$ before the remaining terms of the Hamiltonian are turned on. Here, $P(^{2S+1}\Lambda_u\, 0^\pm)$ denotes the projector onto the respective state. In order to describe the hyperfine and Zeeman structure for a given vibrational level v' (with symmetry $^{2S'+1}\Lambda'_{u'}\, 0^\pm$), the influence of all surrounding vibronic levels for each symmetry is mimicked by a single, effective energy value $E_{\Lambda,S,0^\pm}$. As an example, let us assume that we want to describe the hyperfine and Zeeman structure of the vibrational level $v_b = 75$ of the $b$ state. As can be seen in Fig. 1(b), $v_b = 75$ is surrounded by several $v_A$ levels in its proximity, with $v_A = 67$, 68 and 69 being the closest ones. All these $v_A$ levels are replaced by a single effective vibrational level with energy $E_{\Sigma,0,0^+}$ in our model.

The second term of Eq. (1) is the spin-orbit interaction

$$H_{\rm SO} = C_{\rm SO}(\vec{S}_1 \cdot \vec{L}_1 + \vec{S}_2 \cdot \vec{L}_2), \tag{3}$$

which couples spin $\vec{S}_i$ and orbital angular momentum $\vec{L}_i$ of electron $i$. Here, $C_{\rm SO}$ denotes the spin-orbit parameter being the corresponding atomic value divided by two because we have only 50% probability for each electron to be in the $p$-orbital. From the atomic fine structure in $^{87}$Rb (see, e.g., [34]) one obtains $C_{\rm SO} = \left[E(5^2P_{3/2}) - E(5^2P_{1/2})\right]/(3\hbar^2) = 2374\,\text{GHz} \times h/\hbar^2$. We use this value of $C_{\rm SO}$ for the spin-orbit interaction between $(2)^3\Sigma_u^+$ and $b^3\Pi_u\ 0^-$. These states are separated by about $5000\,\text{cm}^{-1}$ [cf. Fig. 1(a)]. The corresponding level repulsion shifts the $b^3\Pi_u\ 0^-$ component to lower energies by several tens of GHz $\times\, h$ compared to the situation, when spin-orbit interaction is ignored. For the spin-orbit coupling between $A^1\Sigma_u^+$ and $b^3\Pi_u\ 0^+$, we additionally take into account the overlap integral of the relevant vibrational wave functions, which is typically $\sim 0.1$ for states of the considered frequency range ($9360 - 9600\,\text{cm}^{-1}$). The spin-orbit interaction is responsible for the frequency splitting $\Delta$ between the $0^-$ and $0^+$ components of $b^3\Pi_u$ and the mixing of the $A$ and $b$ state which is expressed in terms of the admixing parameter $p_b$. It turns out that $\Delta$ and $p_b$ are the two quantities, which essentially determine the hyperfine and Zeeman structure of a vibrational state. By fine tuning $\Delta$ and $p_b$ in our model we can describe the observed spectra. For practical purposes, we vary neither $C_{\rm SO}$ nor the value of the overlap integral ($= 0.1$), instead we use the term energies of the relevant uncoupled states in $H_{\rm Diag}$. Concretely, we adjust $p_b$ by setting the separation between $A^1\Sigma_u^+$ and $b^3\Pi_u$, while the size of $\Delta$ is adjusted by shifting the term energy of the $0^-$ level relative to the $0^+$ level.

The third term of Eq. (1),

$$H_{\rm R} = B_{\rm v}\vec{R}^2, \tag{4}$$

describes the rotation of the atom pair. According to the calculations of Drozdova *et al.* [28], the rotational constant $B_{\rm v}$ is about $0.54\,\text{GHz} \times h/\hbar^2$ for the $A - b$ states with dominant $b$ character and vibrational quantum numbers $\text{v}_b \sim 70 - 80$ of $^{87}\text{Rb}_2$. The quantum number of angular momentum $\vec{R}$ appearing in the atom pair basis determines the total parity of the molecular state according to $-(-1)^R$. But $R$ is not a good quantum number for the molecular eigenstates since $\vec{R}^2$ does not commute with $H_{\rm Diag}$.

Next, we consider the hyperfine interaction $H_{\rm HF}$. As mentioned in [4], the Fermi contact term is in general sufficient to characterize the hyperfine interaction of alkali-metal dimers. We use

$$H_{\rm HF} = b_{\rm F}(\vec{S} \cdot \vec{I}), \tag{5}$$

with $\vec{S} = \vec{S}_1 + \vec{S}_2$ and $\vec{I} = \vec{I}_1 + \vec{I}_2$. According to Refs. [4, 35, 36], the Fermi contact parameter $b_{\rm F}$ for an atom pair $(s+p)$ is $b_{\rm F} \sim A_{\rm HF,atom}/4$, where $A_{\rm HF,atom} = 3.417\,\text{GHz} \times h/\hbar^2$ denotes the atomic hyperfine parameter for the $5S_{1/2}$ level of $^{87}$Rb [37]. We note that this Ansatz is formally identical to the atomic hyperfine interaction of a ground state electron (i.e., $s$ orbital) with its local nuclear spin $I_i$. The factor $1/4$ normalizes the interaction because at any instant in time only one of the two electrons (i.e., the $s$ electron) interacts with only one of the two nuclei. We fix the Fermi contact parameter to be $b_{\rm F} = A_{\rm HF,atom}/4$ but note that deviations of up to 20% from this approximation have been observed, e.g. for $Na_2$ [36]. We do not use $b_F$ as a free fit parameter in our model because it is strongly correlated with the unknown frequency splitting $\Delta$ (see section 6). Thus, any uncertainty in $b_F$ directly translates into an uncertainty of $\Delta$. Furthermore, by using the ansatz of Eq. (5) we neglect the nondiagonal part of the hyperfine interaction with respect to $S$ and $I$ and thus there is no mixing of $u/g$ symmetry. However, this approximation should be valid as the energy spacing between possibly coupled $u/g$ states is significantly larger than the $0^-/0^+$ spacing considered in this work.

The last term of Eq. (1) characterizes the Zeeman interaction in a homogeneous magnetic field of strength $B$ in $z$ direction

$$H_Z = \mu_B [g_L L_z + g_S S_z + g_I I_z] B, \tag{6}$$

with $\mu_B$ being Bohr's magneton. Here, we consider the Zeeman interaction due to the orbital angular momenta of the electrons, the electronic spins as well as the nuclear spins, where $g_L = 1$, $g_S = 2.002319$ and $g_I = -0.000995$ for $^{87}$Rb [37] are the corresponding $g$-factors.

The matrix elements are calculated in an uncoupled atom pair basis, being a properly antisymmetrized product of eigenstates of all needed angular momenta and their projection on the space-fixed axis $z$, and the nuclear positions at both ends of the rotator axis. Table 3 gives an overview of the range of quantum numbers for the $\Omega = 0$ states of $b^3\Pi_u$, which are needed to setup the matrix. The total molecular angular momentum $J$ ($\vec{J} = \vec{R} + \vec{L} + \vec{S}$, i.e., without nuclear spins) is a fairly good quantum number, because the hyperfine and Zeeman interaction is small compared to the other interactions.

## 6. Model calculations and interpretation of measured data

In the following we use the model introduced in the previous section to calculate the Zeeman and hyperfine structure for an $A-b$ bound state as a function of the frequency splitting $\Delta = [E(0^-, J=0) - E(0^+, J=1)]/h$ of the $0^\pm$ components of $b^3\Pi_u$‡, the degree of mixture $p_b$ between the $A$ and $b$ states, as well as the magnetic field $B$. We define $\Delta$ to be the splitting between $0^\pm$ after diagonalization of the Hamiltonian $H$ of Eq. (1). In order to keep the discussion simple, we restrict ourselves to the range of quantum numbers and parameters directly related to our experiments. As explained in section 3.2, starting from Feshbach molecules we can only optically excite $A-b$ bound levels through the $A^1\Sigma_u^+$ $0^+$ component with angular momentum $J = 1$ and negative parity. Furthermore, we want to point out that the eigenstates of the Hamiltonian of Eq. (1) are eigenstates of the total nuclear spin $I$. Thus, we can restrict ourselves to bound states with $I = 2$, being equal to the value of the singlet component of the Feshbach state applying the electric dipole selection rule $\Delta I = 0$.

The diagonal Zeeman and hyperfine interactions of the states $0^\pm$ are negligible compared to our measurement uncertainty. However, both the Zeeman interaction $H_Z$ and the hyperfine interaction $H_{HF}$ couple $0^+$ and $0^-$ within $b^3\Pi_u$. In particular, ($J = 1$, $I = 2$) of $0^+$ couples to ($J = 0, I = 2$) and ($J = 2, I = 2$) of $0^-$ (see table 3). This leads to mixing, i.e., to the creation of eigenstates with a net electronic magnetic moment and thus to Zeeman and hyperfine splittings. Interestingly, hyperfine and Zeeman interaction amplify the line splittings in a cooperative way because they have matrix elements for the coupling between $0^+$ and $0^-$ similar in magnitude and equal in sign. Hence, if Zeeman and hyperfine interaction are of the same strength their combined effect increases the line spacings not only by a factor of two but by a factor of four. Our experiments are indeed close to this regime for the selected magnetic field of about 1000 G.

‡ The different $J$ values are required for obtaining the same parity for the mixed states. In addition to $J = 0$, also the level $J = 2$ of $0^-$ will couple to $J = 1$ of $0^+$. The rotational energy splitting between $J = 0$ and 2 of $0^-$ is determined by Eq. (4).

**Table 3.** Overview of the range of quantum numbers for states $0^+$ and $0^-$ of $b^3\Pi_u$ with low angular momentum $J$. The $+/-$ columns provide the total parity whereas $I$ represents the total nuclear spin and $R$ is the atom pair rotation. Note that the quantum numbers $I$ and parity alternate with $J$. This behavior is also found for the even and odd values of $R$. Furthermore, the molecular rotation increases with $J$.

| | $0^+$ | | | $0^-$ | | |
|---|---|---|---|---|---|---|
| $J$ | $I$ | $+/-$ | $R$ | $I$ | $+/-$ | $R$ |
| 0 | 1, 3 | $+$ | 1 | 0, 2 | $-$ | 0, 2 |
| 1 | 0, 2 | $-$ | 0, 2 | 1, 3 | $+$ | 1, 3 |
| 2 | 1, 3 | $+$ | 1, 3 | 0, 2 | $-$ | 0, 2, 4 |
| 3 | 0, 2 | $-$ | 2, 4 | 1, 3 | $+$ | 1, 3, 5 |
| 4 | 1, 3 | $+$ | 3, 5 | 0, 2 | $-$ | 2, 4, 6 |
| $\vdots$ | $\vdots$ | $\vdots$ | $\vdots$ | $\vdots$ | $\vdots$ | $\vdots$ |

The Zeeman and hyperfine splitting crucially depends on the frequency spacing $\Delta$ between the levels $0^-$ and $0^+$ of $b^3\Pi_u$. Using standard perturbation theory, the splitting is estimated to be proportional to $\langle H_{\mathrm{Z}} + H_{\mathrm{HF}} \rangle^2/\Delta$. We recall that the spin-orbit couplings to $(2)^3\Sigma_u^+$ and $A^1\Sigma_u^+$ generate the spacing $\Delta$ between the levels $0^-$ and $0^+$ of $b^3\Pi_u$ in the restricted Hilbert space.

Figure 4 depicts results of our model calculations for a vibrational level of $A - b$ with 80% triplet ($b$) and 20% singlet ($A$) character. The frequency positions of levels $|F', m_F'\rangle$ are shown as a function of $\Delta$ for two magnetic fields, $B = 0$ G and 1000 G. Although the total angular momentum $F'$ is generally not a good quantum number anymore at higher magnetic fields, we refer to the molecular levels by the correlated value of $F'$ at $B = 0$ G. In Fig. 4 only those six states $|F', m_F'\rangle$ are plotted that, according to the dipole selection rules, are accessible by our spectroscopy experiment. As the Feshbach molecule has $F = 2$, $m_F = 2$, levels with the quantum numbers $m_F' = 2$ ($m_F' = 1, 3$) can be observed via $\pi$ ($\sigma$) transitions. We choose the state $|F' = 2, m_F' = 2\rangle$ as energy reference (i.e., $\delta = 0$), because it corresponds to the strong $\pi$ resonance in Figs. 2 and 3. This allows for convenient comparison of our calculations to the measured spectra. Note, at $B = 0$ G, no $m_F$ splitting can occur and therefore only three different curves are discernible in Fig. 4(a).

In our calculations, $\Delta$ is set by adjusting the initial energy spacing between the $0^+$ and $0^-$ components of the $b$ state in $H_{\mathrm{Diag}}$ [see Eq. (2)]. To a good approximation, within the frequency ranges considered in Fig. 4 (i.e., $|\Delta| = 16 - 129\,\mathrm{GHz}$) the level splittings increase inversely with $\Delta$, just as expected from perturbation theory. This can directly be seen in Fig. 7 of the Appendix. For $B = 1000\,\mathrm{G}$ and a small $|\Delta|$ of $16\,\mathrm{GHz}$, the overall spreading of the levels reaches more than $200\,\mathrm{MHz}$, whereas for a large $|\Delta| \geq 60\,\mathrm{GHz}$ it is on the order of a few tens of MHz or less. Furthermore, the ordering of the energy levels is inverted, when the sign of $\Delta$ changes. We point out that some level spacings are smaller than the expected linewidths and therefore cannot be resolved in the experiment. The widths of the levels are mainly determined by those of the $A^1\Sigma_u^+$ state ($\sim 12\,\mathrm{MHz}$) and the admixing parameter $p_A$, because the width of the pure $b^3\Pi_u$ state is orders of magnitude smaller compared to the one of $A^1\Sigma_u^+$. The basic structure of the calculated levels (see Fig. 4) and the level widths let us expect to resolve three resonance features which agrees well with our observations shown in Fig. 3.

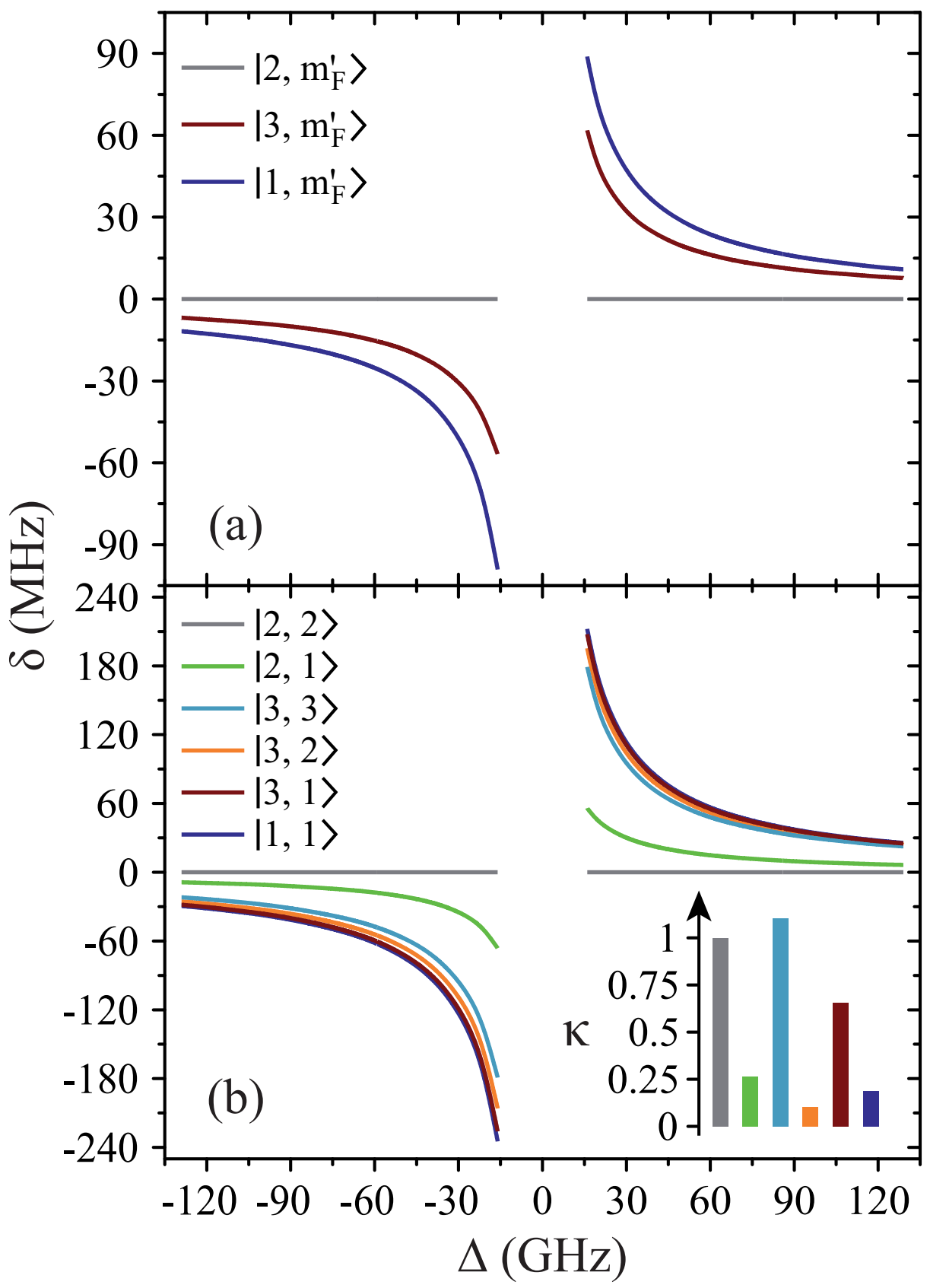

**Figure 4.** Hyperfine level structure for a vibrational state $\mathrm{v}_b$ with $p_b = 80\%$ for two magnetic fields, $B = 0\,\mathrm{G}$ (a) and $1000\,\mathrm{G}$ (b). Shown are the frequency positions $\delta$ of the levels $|F', m_F'\rangle$ relative to the state $|2,2\rangle$, as a function of $\Delta$. The inset in (b) gives the relative strengths $\kappa$ for the optical dipole transitions from the Feshbach state towards the levels $|F', m_F'\rangle$ of $\mathrm{v}_b$ at $B = 1000\,\mathrm{G}$. For $|2,2\rangle$ we set $\kappa = 1$. For convenience, we have plotted the same data in Fig. 7 in terms of $1/\Delta$. This makes it easier to read off the line splittings for small $|\Delta|$.

We now want to assign the experimentally observed resonances to distinct transitions. Besides considering the line positions we also take into account the strength of the lines. For this purpose, we calculate the dipole matrix elements $M_{\mathrm{FS},(F',m_F')}$ from the initial Feshbach state $|\mathrm{FS}\rangle$ to the final levels $|F',m_F'\rangle$ of the mixed $b^3\Pi_u$ state. As already mentioned, we only have to consider the singlet component of both levels. The inset in Fig. 4(b) shows the relative transition strengths $\kappa = M^2_{\mathrm{FS},(F',m_F')}/M^2_{\mathrm{FS},(2,2)}$ at $B = 1000\,\mathrm{G}$. We can roughly group the six transitions into three strong ones (by $\sigma$ light towards final states $|3,1\rangle$ and $|3,3\rangle$, by $\pi$ light towards $|2,2\rangle$) and three weak transitions ($\sigma$: towards $|1,1\rangle$ and $|2,1\rangle$ and $\pi$: towards $|3,2\rangle$). Our observed spectra always exhibit only one strong line for $\pi$-polarized light (cf. Fig. 3). It is therefore assigned to $|2,2\rangle$ and used as reference level. The calculations predict a second resonance for $\pi$ polarization which should be about one order of magnitude weaker. However, we did not unambiguously observe this line since its signal is easily drowned by the overlaying strong $\sigma$ lines if the achieved light polarization is not sufficiently pure.

In the following, the $\sigma$ transitions are discussed in detail using the results shown in Fig. 4(b). Although the transition towards $|2,1\rangle$ is weak, we should be able to clearly observe it, since the level $|2,1\rangle$ is well separated from all other levels. The three remaining transitions (towards $|1,1\rangle$, $|3,1\rangle$ and $|3,3\rangle$), however, are quite close to each other. Especially for larger values of $|\Delta|$ ($\gtrsim 40\,\mathrm{GHz}$) these levels cannot be resolved and only a single resonance should be visible. Among the three transitions, the one towards $|3,3\rangle$ is most dominant. For low values of $|\Delta|$ ($< 40\,\mathrm{GHz}$) the strong $|3,3\rangle$ line splits clearly from those corresponding to $|1,1\rangle$ and $|3,1\rangle$ which both barely separate. This explains why our observed spectra for $\sigma$-polarized light in Fig. 3 exhibit at most three resonance features. At this stage we have shown that the experimental data can be qualitatively explained by our theoretical model and that we already can assign quantum numbers to the measured resonance lines.

Now, we want to carry out a more quantitative comparison of the measured line splittings in Fig. 3 with the model predictions. For this, we study the dependence of the energy level structure on the admixing parameter $p_b$, i.e., the percentage of the $b^3\Pi_u$ potential in the vibrational state $\mathrm{v}_b$. In the simulations, we set $p_b$ by adjusting the term energy of the bare $A^1\Sigma_u^+$ state in Eq. (2). Results for $\Delta = +59\,\mathrm{GHz}$ are shown in Fig. 5. To good approximation, within the investigated range from $p_b = 65\%$ to $95\%$ the level frequencies depend linearly on $p_b$. This makes sense as the discussed hyperfine and Zeeman interaction only appears within the $b^3\Pi_u$ ($\Omega = 0$) state.

In order to carry out the quantitative comparison for each spectrum $\mathrm{v}_b$, we individually fix the splitting of the bare $A^1\Sigma_u^+$ $0^+$ and $b^3\Pi_u$ $0^+$ levels such that the admixing parameter $p_b$ equals to its literature value [1, 28], as listed in table 2. Afterwards, we fit our model to the measured spectrum by adjusting a single parameter, the effective term energy of $b^3\Pi_u$ $0^-$ in $H_{\mathrm{Diag}}$ and thus the splitting $\Delta$ between the $0^-$ and $0^+$ components of the $b$ state. All other parameters of the model are kept at the values given in section 5. For the fit, we ignore the states $|3,2\rangle$ and $|1,1\rangle$ since they cannot be experimentally resolved [see inset of Fig. 4(b)]. The resulting spectral positions of the hyperfine levels are shown in Fig. 3 as vertical lines together with the measured spectra. The agreement is quite satisfactory as the experimental and calculated line positions do not differ by more than a few MHz. Our fit results for $\Delta$, i.e., the splitting of the $0^\pm$ states after diagonalizing the Hamiltonian $H$ of Eq. (1), are listed in Fig. 3 as well as in table 2. We obtain values ranging from $\Delta = -19.7\,\mathrm{GHz}$ to $104.5\,\mathrm{GHz}$. The error boundaries for $\Delta$ in table 2 are estimated by simulations shifting the resonance frequency $\delta$ of the strong $|F' = 3, m'_F = 1\rangle$ line by $\pm 5\,\text{MHz}$ relative to the reference $|2,2\rangle$. Such an approach is reasonable as the frequency stability in our measurements is $\pm(2-5)\text{MHz}$. In addition to these error boundaries there exists a further uncertainty in $\Delta$, since the precise value of the hyperfine interaction parameter $b_{\text{F}}$ is unknown. As already discussed in section 5, we expect that our adopted value for $b_{\text{F}}$ possibly deviates by up to 20% from its real value. According to the simple estimation for the hyperfine and Zeeman splittings ($\propto \langle H_{\text{Z}} + H_{\text{HF}}\rangle^2/\Delta$), such a change in $b_{\text{F}}$ leads to a similar change of $|\Delta|$ from the fit. In our data analysis, we have verified that the fits to the measured spectra remain of similar quality when we vary $b_{\text{F}}$ by a few tens of %.

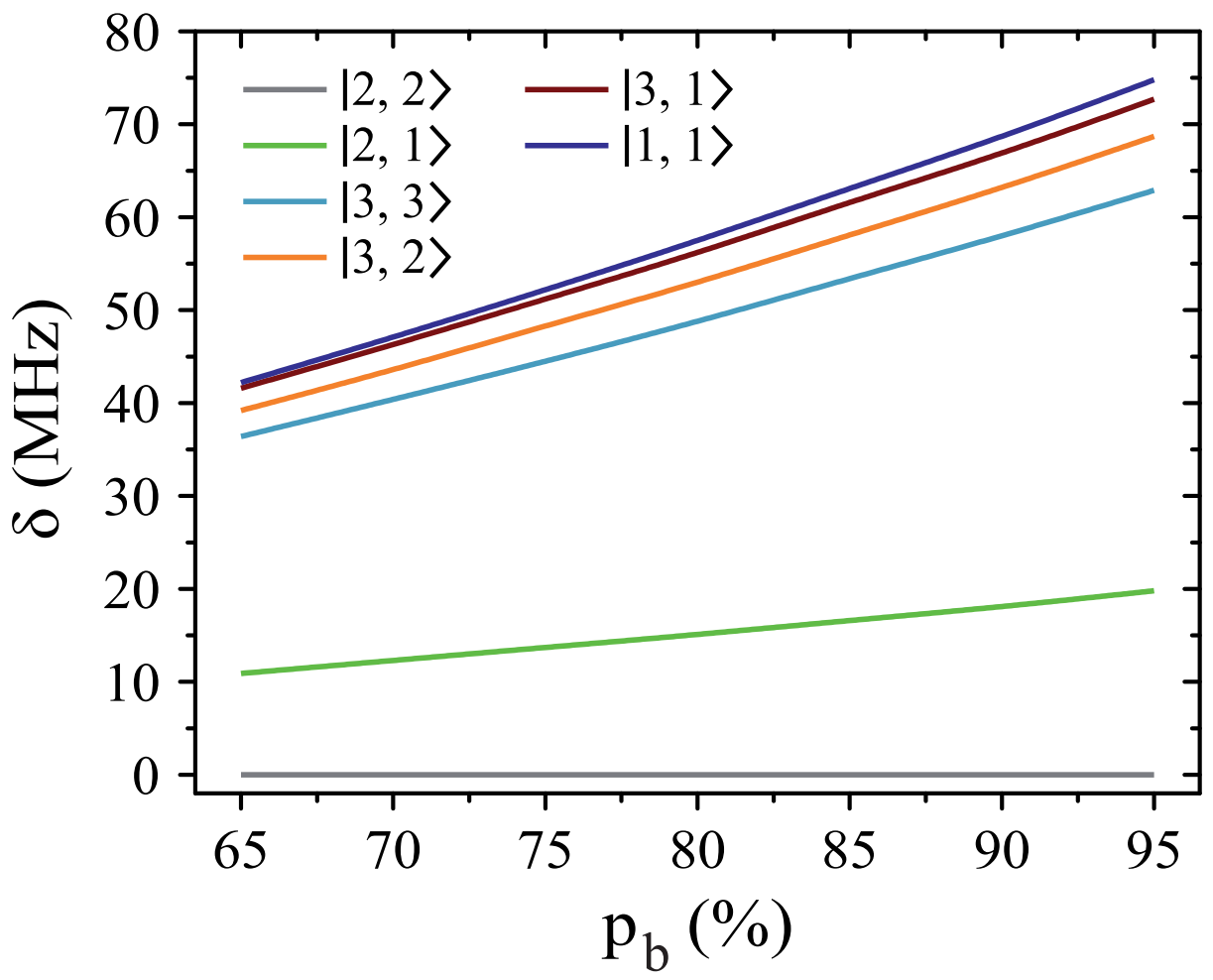

**Figure 5.** Dependence of the hyperfine and Zeeman structure on the admixing parameter $p_b$ for a fixed value of $\Delta = +59\,\text{GHz}$ and a magnetic field of $B = 1000\,\text{G}$. The frequency $\delta$ is given relative to the $|2,2\rangle$ level.

Next, we investigate the Zeeman and hyperfine structure as a function of the magnetic field. Figures 6(a) and (b) depict the model calculations for the vibrational levels $\text{v}_b = 75$ and 73, respectively, using the values of $p_b$ and $\Delta$ given in table 2. For both plots the admixing parameters are similar ($p_b \approx 80\%$), while the respective frequency spacings $\Delta$ have different signs and magnitude. We show all the levels accessible in our spectroscopy together with the experimentally derived levels at $B = 999.9\,\text{G}$. Here, the frequency reference is represented by the level $|2,2\rangle$ at this magnetic field. We present in Fig. 6(c) the full hyperfine and Zeeman structure of a single state $\text{v}_b$ for $\Delta = +14.0\,\text{GHz}$, $p_b = 80.00\%$. This graph reveals particularly well the transition from the linear Zeeman effect to a quadratic behavior above a few hundreds of gauss and the enhancement of the splitting by the cooperative effect between Zeeman and hyperfine interaction.

Finally, we want to give a quantitative interpretation of the spectrum corresponding to the vibrational level $\text{v}_A = 72$ of $A^1\Sigma_u^+$ [see Fig. 1(b)], which has a large $b$ state admixture ($p_b = 39.36\%$) due to the strong coupling of $\text{v}_A = 72$ to $\text{v}_b = 78$ as these states are fairly close to each other. The separation is only 89.2 GHz according to tables 1 and 2. For $\text{v}_b = 78$, our model determines a spacing of $\Delta = 36.4^{+3.4}_{-2.9}$ GHz between its $0^+$ and $0^-$ components. Consequently, the $b^3\Pi_u$ $0^-$ state is only separated by $-52.8^{+3.4}_{-2.9}$ GHz from the $A^1\Sigma_u^+$ $0^+$ state. From this, we can predict the hyperfine structure for $\mathrm{v}_A = 72$ with our model. The results are shown in Fig. 1(b). As can be seen, the calculated and measured resonances agree well, which nicely confirms the consistency of our model.

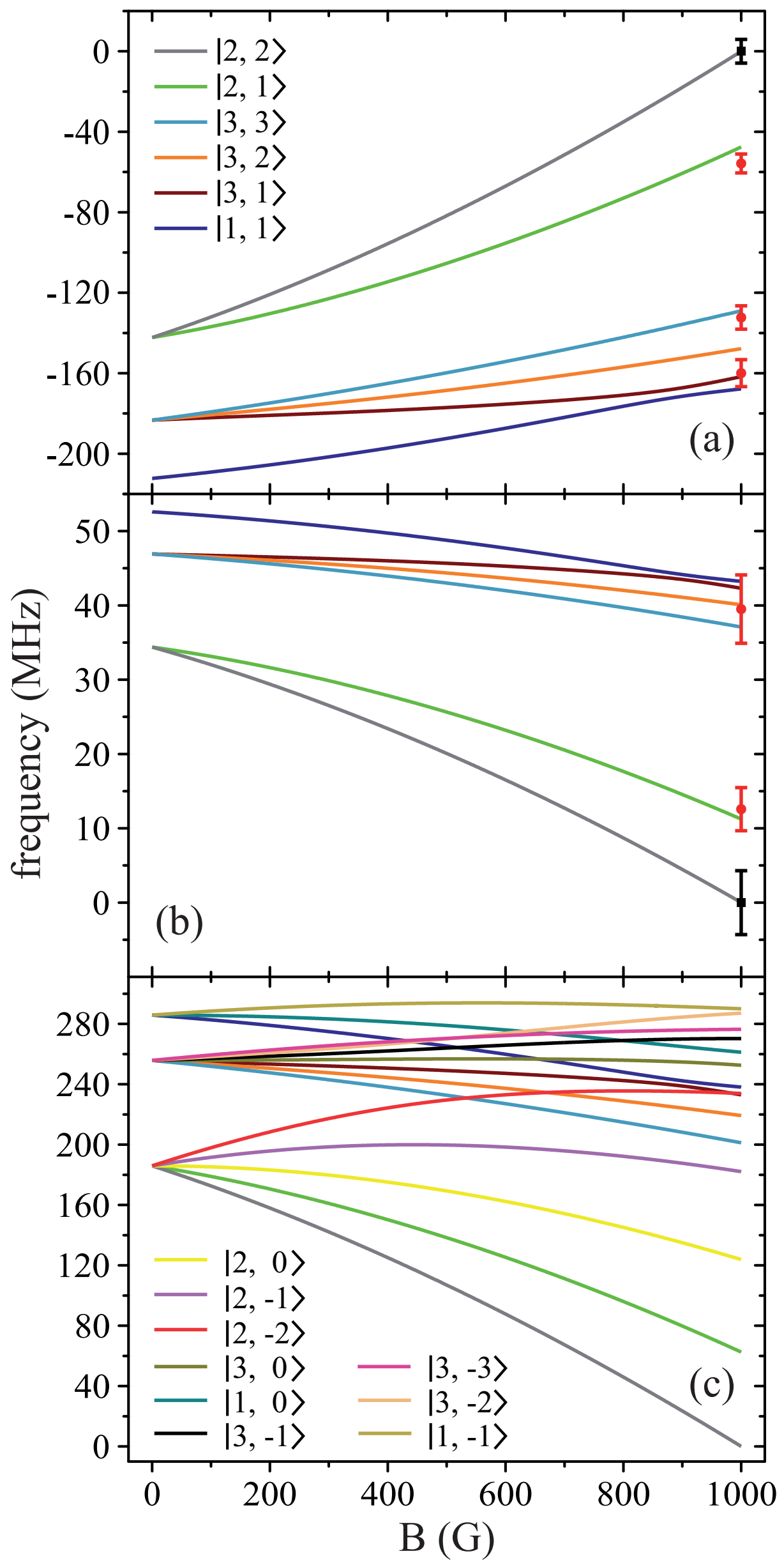

**Figure 6.** Zeeman structure of the hyperfine levels $|F', m'_F\rangle$ as a function of the magnetic field $B$. The frequency is referenced to the position of level $|2,2\rangle$ at $B = 999.9\,\mathrm{G}$. (a) Simulations for $\Delta = -19.7\,\mathrm{GHz}$, $p_b = 73.80\%$, (b) for $\Delta = +81.8\,\mathrm{GHz}$, $p_b = 82.70\%$ and (c) for $\Delta = +14.0\,\mathrm{GHz}$, $p_b = 80.00\%$. The black square (red dot) plot symbols indicate the experimentally observed resonances obtained with $\pi$-polarized ($\sigma$-polarized) light corresponding to $\mathrm{v}_b = 75$ (a) and 73 (b). The error bars represent the measured transition linewidths (FWHM) determined from our fits to the data [cf. Fig. 3(b) and (e)]. In (c), all hyperfine energy levels of the vibrational state $\mathrm{v}_b$ ($J = 1$, $I = 2$) are plotted, where the color code is extended for the additional levels compared to (a) and (b).

**Table 4.** Energy levels of the component $\Omega = 0$ of $A^1\Sigma_u^+$ and of $b^3\Pi_u$ without (w/o SO) and with (w SO) spin-orbit coupling. Column 3 shows the calculated energies for $0^+$ of the $A$ and $b$ states for $J = 1$, whereas column 4 contains only energy levels for $0^-$ and $J = 0$. In column 5 the corresponding values for $0^+$ and $J = 1$ are provided, obtained with the spin-orbit interaction between states $A$ and $b$. Column 6 reports the splittings $[E(0^-) - E(0^+)]/h$ calculated for the restricted system $A - b$ and denoted by $\Delta_{\mathrm{A-b}}$, while column 7 lists the results derived from our experiments. (For details see text.)

| state | $\mathrm{v}_A$, $\mathrm{v}_b$ | $0^+, J = 1$ (w/o SO) $[hc \times \mathrm{cm}^{-1}]$ | $0^-, J = 0$ (w/o SO) $[hc \times \mathrm{cm}^{-1}]$ | $0^+, J = 1$ (w SO) $[hc \times \mathrm{cm}^{-1}]$ | $\Delta_{\mathrm{A-b}}$ [GHz] | $\Delta_{\mathrm{obs}}$ [GHz] | $\Delta_{\mathrm{obs}} - \Delta_{\mathrm{A-b}}$ [GHz] |
|---|---|---|---|---|---|---|---|
| $A^1\Sigma_u^+$ | 65 | 9352.911 | | 9352.078 | 534.3 | | |
| $b^3\Pi_u$ | 73 | 9369.939 | 9369.901 | 9368.758 | 34.3 | 81.8 | $47.5^{+10.6}_{-8.5}$ |
| $A^1\Sigma_u^+$ | 66 | 9388.026 | | 9388.005 | -542.7 | | |
| $b^3\Pi_u$ | 74 | 9414.679 | 9414.641 | 9412.519 | 63.6 | 104.5 | $40.9^{+23.9}_{-16.4}$ |
| $A^1\Sigma_u^+$ | 67 | 9423.000 | | 9423.589 | -268.3 | | |
| $A^1\Sigma_u^+$ | 68 | 9457.834 | | 9454.571 | 137.9 | | |
| $b^3\Pi_u$ | 75 | 9459.210 | 9459.172 | 9460.874 | -51.0 | -19.7 | $31.3^{+0.6}_{-0.6}$ |
| $A^1\Sigma_u^+$ | 69 | 9492.527 | | 9491.049 | 373.0 | | |
| $b^3\Pi_u$ | 76 | 9503.529 | 9503.492 | 9503.516 | -0.7 | 40.0 | $40.7^{+2.2}_{-2.0}$ |
| $A^1\Sigma_u^+$ | 70 | 9527.078 | | 9525.742 | 655.2 | | |
| $b^3\Pi_u$ | 77 | 9547.635 | 9547.598 | 9547.629 | -0.9 | | |
| $A^1\Sigma_u^+$ | 71 | 9561.487 | | 9560.041 | -373.0 | | |
| $b^3\Pi_u$ | 78 | 9591.525 | 9591.488 | 9591.479 | 0.3 | 36.4 | $36.1^{+3.4}_{-2.9}$ |
| $A^1\Sigma_u^+$ | 72 | 9595.754 | | 9594.454 | -88.9 | -52.8 | $36.1^{+3.4}_{-2.9}$ |
| $A^1\Sigma_u^+$ | 73 | 9629.979 | | 9627.657 | 224.9 | | |

## 7. Splitting between $0^+$ and $0^-$ components in a potential scheme

In the previous section, we have determined the effective splitting $\Delta = [E(0^-, J = 0) - E(0^+, J = 1)]/h$ for the state $b^3\Pi_u$ using our simple model without vibrational degree of freedom. Here, we compare the obtained results to those of coupled channel calculations with a potential scheme, i.e., including the full dynamics of the relative motion within the atom pair. As a first step we follow Ref. [1] and therefore restrict the calculations to the $A - b$ system, such that the spin-orbit interaction and the molecular rotation only couple the $A$ and $b$ states.

Thus, the influence of the spin-orbit coupling of $b^3\Pi_u$ $0^-$ to $(2)^3\Sigma_u^+$ $0^-$ is not yet considered. The term values for the $A-b$ complex are listed in table 4. For the calculations we take the model potentials and spin-orbit functions reported in Ref. [1]. Columns 3 and 4 show the term values for $0^+$, $J=1$ and $0^-$, $J=0$ in the absence of spin-orbit coupling, respectively. At this stage, the $0^+$ and $0^-$ components of the $b^3\Pi_u$ state are only split due to rotation. Column 5 lists the term energies for $0^+$, $J=1$ if the spin-orbit coupling is included. These are the same values as given in tables 1 and 2. Column 6 provides the energy difference of each $0^+$ state in column 4 relative to the closest $0^-$ state in column 3. This quantity, which we call $\Delta_{\mathrm{A-b}}$, is the prediction for the $0^\pm$ splitting within the $A-b$ system. Noticeably, $\Delta_{\mathrm{A-b}}$ varies strongly for different vibrational levels, which results from the fact, that the uncoupled vibrational ladders $\mathrm{v}_A$ and $\mathrm{v}_b$ of $A^1\Sigma_u^+$ and $b^3\Pi_u$ are interwoven with different spacings. Thus, their repulsion due to the interaction of different vibrational levels is fairly irregular. Column 7, labeled $\Delta_{\mathrm{obs}}$, recalls the findings for $\Delta$ inferred from our measurements. There is disagreement between the measurements and the results of the restricted coupled channel model. All values $\Delta_{\mathrm{obs}}$ are significantly larger (more positive) than $\Delta_{\mathrm{A-b}}$. However, it is striking that the differences between $\Delta_{\mathrm{obs}}$ and $\Delta_{\mathrm{A-b}}$, are all close to 40 GHz. The actual values are given in column 8, where the uncertainties are taken from $\Delta_{\mathrm{obs}}$ (see, e.g., table 2). Indeed, all differences (except for $\mathrm{v}_b = 75$) are equal within the given uncertainties. Could the spin-orbit coupling to state $(2)^3\Sigma_u^+$ be responsible for this discrepancy of $\approx 40$ GHz? No, it cannot. Indeed, the $(2)^3\Sigma_u^+$ is far up in energy [cf. Fig. 1(a)] and therefore all vibrational levels of $b^3\Pi_u$ $0^-$ experience an almost constant shift, but it has the wrong sign to explain the observations. If *ab initio* calculations for the $(2)^3\Sigma_u^+$ potential energy curve and the corresponding spin-orbit interaction are applied, we obtain quantitatively that the $0^-$ component of the $b^3\Pi_u$ state is shifted by about $50\,\mathrm{GHz}\times h$ to lower energies. This increases the deviation of already 40 GHz to about 90 GHz, much beyond experimental uncertainties.

Because the energetic order of the $0^+$ components are accurately determined from former spectroscopic work [1], our observation directs clearly to the spin-orbit energy of the $0^-$ component of state $b^3\Pi_u$. Interestingly, the spin-orbit functions of $b^3\Pi_u$ presented in Fig. 6(a) of [1] are different for $\Omega=2$ and $\Omega=0$ by about $1.8\,\mathrm{cm}^{-1}$ as derived from the reported amplitude $D_e^{SO}$ of the spin-orbit function in Table V of Ref. [1]. This difference indicates that $b^3\Pi_u$ is no longer a pure spin-orbit multiplet of a $\Pi$ state. If we consider that a similar deviation of the multiplet structure also exists for $0^-$, then a decrease of about $2.6\,\mathrm{cm}^{-1}$ compared to the spin-orbit function for $0^+$ in [1] would be sufficient to model the $\Delta_{\mathrm{obs}}$ as given in table 4. We believe that this is a plausible explanation of our observation.

Finally, in view of the calculated splittings of $0^-$ and $0^+$ we want to recall the measurements for the vibrational levels $\mathrm{v}_A = 66$ and 69 of $A^1\Sigma_u^+$ (cf. Fig. 2(a) and table 1). The absolute values $|\Delta_{\mathrm{A-b}}|$, even when corrected by the above determined shift of 90 GHz, are very large. Therefore, according to the discussion in section 6, we do not expect significant hyperfine and Zeeman splittings. Indeed, no hyperfine structure was observed in our spectra for $\mathrm{v}_A = 66$ and 69.

## 8. Conclusion

We have investigated the combined hyperfine and Zeeman structure in the spin-orbit coupled $A^1\Sigma_u^+ - b^3\Pi_u$ complex of $^{87}Rb_2$ dimers. We performed spectroscopy of ultracold Feshbach molecules at a magnetic field of 999.9 G and recorded the spectra for several excited vibrational levels with either dominant $A^1\Sigma_u^+$ character ($v_A = 66-70$ and 72) or dominant $b^3\Pi_u$ character ($v_b = 73-76$ and 78). We observe large line splittings of up to 160 MHz and find that the Zeeman and hyperfine structure of the $0^+$ state of $b^3\Pi_u$ varies strongly for different vibrational levels. Using a simple model, where the molecule is treated as a rigid rotor of two neutral atoms, we can explain the level structures as resulting from nondiagonal hyperfine and Zeeman interactions between the $0^+$ and $0^-$ components of $b^3\Pi_u$. The hyperfine and Zeeman interactions act in a cooperative way, which enhances the level splittings. Furthermore, the level splittings depend linearly on the admixture $p_b$ for the components $0^+$ of the complex $b^3\Pi_u$ and $A^1\Sigma_u^+$, and scale inversely with the frequency spacing $\Delta$ between $0^-$ and $0^+$. From fits of our model to the data, we find that $\Delta$ strongly varies in the range of $-53$ to 105 GHz within the interval of studied $v_b$ and $v_A$. Our observed values for $\Delta$ systematically deviate by about 90 GHz from predictions of close-coupled channel calculations of the electronic structure correlated to the atom pair asymptote $5^2S+5^2P$ using empirical potentials for $A^1\Sigma_u^+$ and $b^3\Pi_u$ and *ab initio* results for $(2)^3\Sigma_u^+$. We can eliminate this deviation by introducing a spin-orbit coupling function that is different by 2.6 $\mathrm{cm}^{-1}$ for $0^-$ and $0^+$.

The fact, that we can extract the frequency spacing $\Delta$ from our measurements of the hyperfine and Zeeman structure is a significant result of this work. In ordinary spectroscopy, starting from a ground state molecule only the state $0_u^+$ can be addressed, owing to selection rules. Therefore, very little data on the $\Lambda$-type doubling of $\Omega = 0$ is available, so far. Recently, all multiplet components of the $(1)^3\Pi_g$ state were observed in the photoassociation of ultracold $^{85}$Rb atoms [38]. The colliding $^2S+^2S$ atom pair contains a superposition of states with both symmetries $^3\Sigma_u^+$ ($0^-$ and 1) from which $(1)^3\Pi_g$ ($\Omega = 0^{\pm}, 1, 2$) can be reached by electric dipole transitions. To our knowledge, our method represents the first work to experimentally access $\Delta$ by investigating only one of the $0^{\pm}$ components of $b^3\Pi_u$.

In the simple model only the hyperfine contribution from the $s$ electron was considered. Thus it is unknown, how much the missing contribution by the $p$ electron and the electric quadrupole interaction might influence the determination of the energy splitting between $0^-$ and $0^+$. In this respect and according to the evaluated multiplet structure here, it would be interesting to compare the results of the simple model to full close-coupled channel calculations based on accurate Born-Oppenheimer potential curves including hyperfine and Zeeman interaction with its non-diagonal contributions between the multiplet components. But for this task more experimental data is needed in order to determine the vibrational dependence over a large interval, which allows for deriving $r$-dependent functional forms of the spin-orbit interaction and probably also of the hyperfine interaction. The latter one was clearly needed in an earlier study on the triplet ground state $a^3\Sigma_u^+$ of $Rb_2$ [31]. We want to emphasize that the strength of the present approach is, that the modeling of all observations on hyperfine and Zeeman splittings are concentrated in an effective single parameter, an average value of the spin-orbit function for $0^-$. The hyperfine parameter and the $g$ factors are kept constant at their atomic values. This will be altered if a wide range of vibrational levels is studied.

In general, the gained information is valuable for a detailed, fundamental understanding of molecular hyperfine and Zeeman level structures. Despite the already existing long tradition (see, e.g., the early paper by Freed [33]) there is still great interest in this field of research as documented by recent experimental studies, for instance with respect to $Rb_2$ [30, 39, 40]. Our results can be exploited for an optimized preparation of ultracold deeply-bound ground state molecules either via photoassociation [41] or a stimulated Raman adiabatic transfer (STIRAP) starting from Feshbach molecules.

**Acknowledgments**

The authors would like to thank Anastasia Drozdova and Paul S. Julienne for valuable information and fruitful discussions. This work is funded by the German Research Foundation (DFG). E.T. gratefully acknowledges support from the Minister of Science and Culture of Lower Saxony, Germany, by providing a Niedersachsenprofessur.

**Appendix**

In figure 7 we show the hyperfine level structure for the parameters of Fig. 4 but as a function of $1/\Delta$. Figure 7 clearly reveals the inverse dependence of the level splittings on $\Delta$ in the investigated range of $|\Delta| = 16 - 129$ GHz. This representation is more convenient for reading off the splitting for small values of $\Delta$. Furthermore, it becomes obvious that the Zeeman effect enhances the splittings since in (b) the frequency separation of the group of levels characterized by $F = 1$ and 3 relative to the group corresponding to $F = 2$ is significantly larger than in (a).

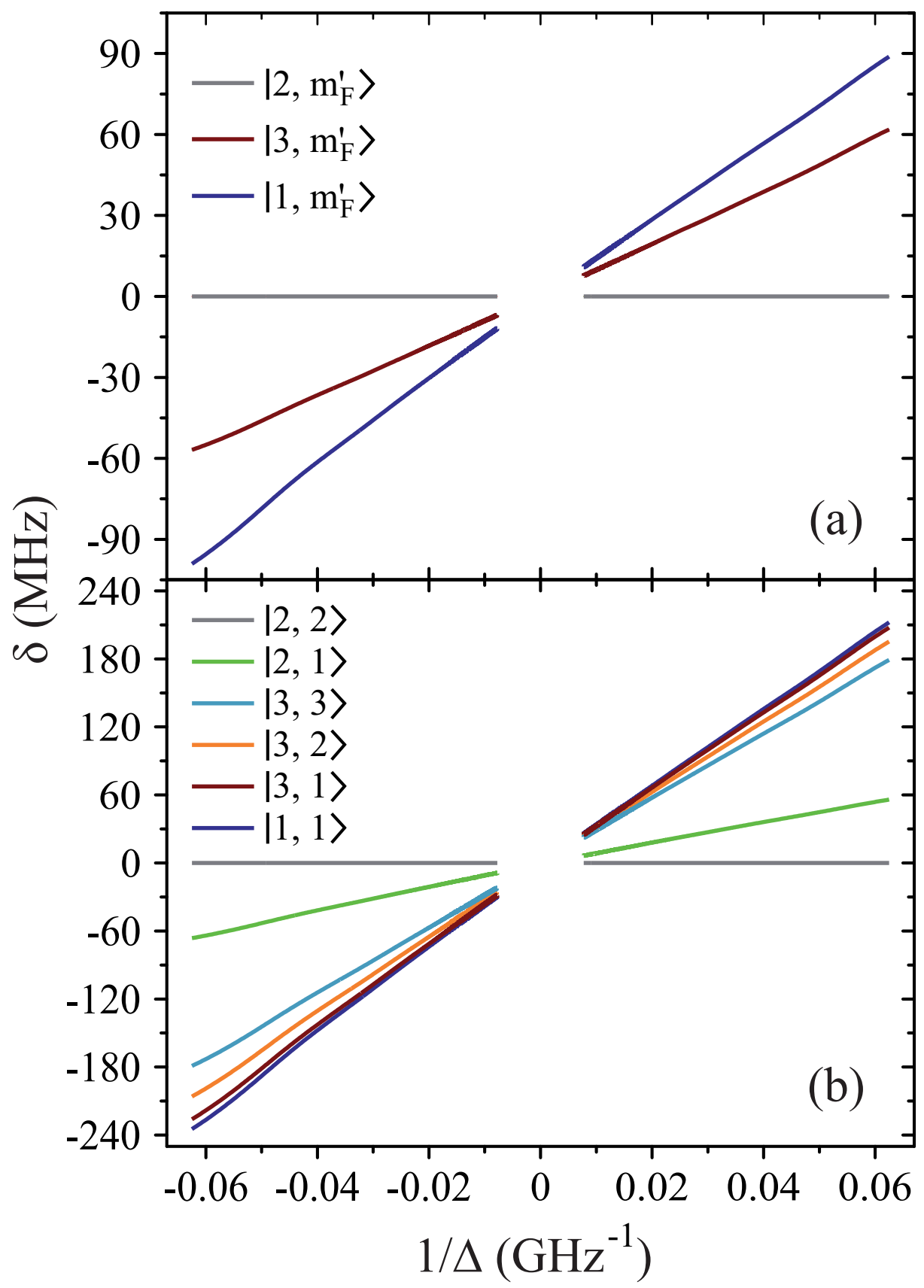

**Figure 7.** Hyperfine level structure for a vibrational state $v_b$ with $p_b = 80\%$ for two magnetic fields, $B = 0\,\text{G}$ (a) and $1000\,\text{G}$ (b). Shown are the frequency positions $\delta$ of the levels $|F', m'_F\rangle$ relative to the state $|2,2\rangle$, as a function of $1/\Delta$.

## References

[1] Drozdova A N, Stolyarov A V, Tamanis M, Ferber R, Crozet P, and Ross A J 2013 Fourier transform spectroscopy and extended deperturbation treatment of the spin-orbit-coupled $A^1\Sigma_u^+$ and $b^3\Pi_u$ states of the $Rb_2$ molecule *Phys. Rev.* A **88** 022504

[2] Salami H, Bergeman T, Beser B, Bai J, Ahmed E H, Kotochigova S, Lyyra A M, Huennekens J, Lisdat C, Stolyarov A V, Dulieu O, Crozet P, and Ross A J 2009 Spectroscopic observations, spin-orbit functions, and coupled-channel deperturbation analysis of data on the $A^1\Sigma_u^+$ and $b^3\Pi_u$ states of $Rb_2$ *Phys. Rev.* A **80** 022515

[3] Amiot C, Dulieu O, and Vergès J 1999 Resolution of the apparent disorder of the $Rb_2$ $A^1\Sigma_u^+$ $(0_u^+)$ and $b^3\Pi_u$ $(0_u^+)$ spectra: A case of fully coupled electronic states *Phys. Rev. Lett.* **83** 2316

[4] Bai J, Ahmed E H, Beser B, Guan Y, Kotochigova S, Lyyra A M, Ashman S, Wolfe C M, Huennekens J, Xie F, Li D, Li L, Tamanis M, Ferber R, Drozdova A, Pazyuk E, Stolyarov A V, Danzl J G, Nägerl H-C, Bouloufa N, Dulieu O, Amiot C, Salami H, and Bergeman T 2011 Global analysis of data on the spin-orbit-coupled $A^1\Sigma_u^+$ and $b^3\Pi_u$ states of $Cs_2$ *Phys. Rev.* A **83** 032514

[5] Danzl J G, Mark M J, Haller E, Gustavsson M, Bouloufa N, Dulieu O, Ritsch H, Hart R, and Nägerl H-C 2009 Precision molecular spectroscopy for ground state transfer of molecular quantum gases *Faraday Discuss.* **142** 283

[6] Qi P, Bai J, Ahmed E, Lyyra A M, Kotochigova S, Ross A J, Effantin C, Zalicki P, Vigué J, Chawla G, Field R W, Whang T-J, Stwalley W C, Knöckel H, Tiemann E, Shang J, Li L, and Bergeman T 2007 New spectroscopic data, spin-orbit functions, and global analyis of data on the $A^1\Sigma_u^+$ and $b^3\Pi_u$ states of

$Na_2$ *J. Chem. Phys.* **127** 044301

[7] Atkinson J B, Becker J, and Demtröder W 1982 Hyperfine structure of the 625 nm band in the $a^3\Pi_u \leftarrow X^1\Sigma_g$ transitions of $Na_2$ *Chem. Phys. Lett.* **87** 128

[8] Falke S, Sherstov I, Tiemann E, and Lisdat C 2006 The $A^1\Sigma_u^+$ state of $K_2$ up to the dissociation limit *J. Chem. Phys.* **125** 224303

[9] Manaa M R, Bai J, Ross A J, Martin F, Crozet P, Lyyra A M, Li L, Amiot C, and Bergeman T 2002 Spin-orbit interactions, new spectral data, and deperturbation of the coupled $b^3\Pi_u$ and $A^1\Sigma_u^+$ states of $K_2$ *J. Chem. Phys.* **117** 11208

[10] Lisdat C, Dulieu O, Knöckel H, and Tiemann E 2001 Inversion analysis of $K_2$ coupled electronic states with the Fourier grid method *Eur. Phys. J.* D **17** 319

[11] Lisdat C, Knöckel H, and Tiemann E 2000 First observation of hyperfine structure in $K_2$ *J. Mol. Spectrosc.* **199** 81

[12] Urbanski K, Antonova S, Yiannopoulou A, Lyyra A M, Li L, and Stwalley W C 1996 All optical triple resonance spectroscopy of the $A^1\Sigma_u^+$ state of $^7Li_2$ *J. Chem. Phys.* **104** 2813

[13] Linton C, Martin F, Russier I, Ross A J, Crozet P, Churassy S, and Bacis R 1996 Observation and analysis of the $A^1\Sigma_u^+$ state of $^6Li_2$ from $v = 0$ to the dissociation limit *J. Mol. Spectrosc.* **175** 340

[14] Docenko O, Tamanis M, Ferber R, Pazyuk E A, Zaitsevskii A, Stolyarov A V, Pashov A, Knöckel H, and Tiemann E 2007 Deperturbation treatment of the $A^1\Sigma^+ - b^3\Pi$ complex of NaRb and prospects for ultracold molecule formation in $X^1\Sigma^+$ ($v = 0$; $J = 0$) *Phys. Rev.* A **75** 042503

[15] Tamanis M, Ferber R, Zaitsevskii A, Pazyuk E A, Stolyarov A V, Chen H, Qi J, Wang H, and Stwalley W C 2002 High resolution spectroscopy and channel-coupling treatment of the $A^1\Sigma^+ - b^3\Pi$ complex of NaRb *J. Chem. Phys.* **117** 7980

[16] Debatin M, Takekoshi T, Rameshan R, Reichsöllner L, Ferlaino F, Grimm R, Vexiau R, Bouloufa N, Dulieu O, and Nägerl H-C 2011 Molecular spectroscopy for ground-state transfer of ultracold RbCs molecules *Phys. Chem. Chem. Phys.* **13** 18926

[17] Docenko O, Tamanis M, Ferber R, Bergeman T, Kotochigova S, Stolyarov A V, de Faria Nogueira A, and Fellows C E 2010 Spectroscopic data, spin-orbit functions, and revised analysis of strong perturbative interactions for the $A^1\Sigma^+$ and $b^3\Pi$ states of RbCs *Phys. Rev.* A **81** 042511

[18] Bergeman T, Fellows C E, Gutterres R F, and Amiot C 2003 Analysis of strongly coupled electronic states in diatomic molecules: Low-lying states of RbCs *Phys. Rev.* A **67** 050501(R)

[19] Kim J-T, Lee Y, Kim B, Wang D, Stwalley W C, Gould P L, and Eyler E E 2011 Spectroscopic analysis of the coupled $1^1\Pi$, $2^3\Sigma^+$ ($\Omega = 0^-, 1$), and $b^3\Pi$ ($\Omega = 0^\pm, 1, 2$) states of the KRb molecule using both ultracold molecules and molecular beam experiments *Phys. Chem. Chem. Phys.* **13** 18755

[20] Zaharova J, Tamanis M, Ferber R, Drozdova A N, Pazyuk E A, and Stolyarov A V 2009 Solution of the fully-mixed-state problem: Direct deperturbation analysis of the $A^1\Sigma^+ - b^3\Pi$ complex in a NaCs dimer *Phys. Rev.* A **79** 012508

[21] Kruzins A, Nikolayeva O, Klincare I, Tamanis M, Ferber R, Pazyuk E A, and Stolyarov A V 2013 Fourier-transform spectroscopy of $(4)^1\Sigma^+ \to A^1\Sigma^+ - b^3\Pi$, $A^1\Sigma^+ - b^3\Pi \to X^1\Sigma^+$, and $(1)^3\Delta_1 \to b^3\Pi_{0^\pm}$ transitions in KCs and deperturbation treatment of $A^1\Sigma^+$ and $b^3\Pi$ states *J. Chem. Phys.* **139** 244301

[22] Kruzins A, Klincare I, Nikolayeva O, Tamanis M, Ferber R, Pazyuk E A, and Stolyarov A V 2010 Fourier-transform spectroscopy and coupled-channels deperturbation treatment of the $A^1\Sigma^+ - b^3\Pi$ complex of KCs *Phys. Rev.* A **81** 042509

[23] Tamanis M, Klincare I, Kruzins A, Nikolayeva O, Ferber R, Pazyuk E A, and Stolyarov A V 2010 Direct excitation of the "dark" $b^3\Pi$ state predicted by deperturbation analysis of the $A^1\Sigma^+ - b^3\Pi$ complex in KCs *Phys. Rev.* A **82** 032506

[24] Ferber R, Pazyuk E A, Stolyarov A V, Zaitsevskii A, Kowalczyk P, Chen H, Wang H, and Stwalley W C 2000 The $c^3\Sigma^+$, $b^3\Pi$, and $a^3\Sigma^+$ states of NaK revisited *J. Chem. Phys.* **112** 5740

[25] Sun H and Huennekens J 1992 Spin-orbit perturbations between the $A(2)^1\Sigma^+$ and $b(1)^3\Pi_0$ states of NaK *J. Chem. Phys.* **97** 4714

[26] Ross A J, Effantin C, d'Incan J, and Barrow R F 1986 Laser-induced fluorescence of NaK: the $b(1)^3\Pi$ state *J. Phys. B: At. Mol. Phys.* **19** 1449

[27] Lozeille J, Fioretti A, Gabbanini C, Huang Y, Pechkis H K, Wang D, Gould P L, Eyler E E, Stwalley W C, Aymar M, and Dulieu O 2006 Detection by two-photon ionization and magnetic trapping of cold $Rb_2$ triplet state molecules *Eur. Phys. J.* D **39** 261

[28] See Supplemental Material of [1] given at http://link.aps.org/supplemental/10.1103/PhysRevA.88.022504.

[29] Thalhammer G, Winkler K, Lang F, Schmid S, Grimm R, and Hecker Denschlag J 2006 Long-lived Feshbach molecules in a three-dimensional optical lattice *Phys. Rev. Lett.* **96** 050402

[30] Takekoshi T, Strauss C, Lang F, Hecker Denschlag J, Lysebo M, and Veseth L 2011 Hyperfine, rotational, and Zeeman structure of the lowest vibrational levels of the $^{87}Rb_2$ $(1)^3\Sigma_g^+$ state *Phys. Rev.* A **83** 062504

[31] Strauss C, Takekoshi T, Lang F, Winkler K, Grimm R, Hecker Denschlag J, and Tiemann E 2010 Hyperfine, rotational, and vibrational structure of the $a^3\Sigma_u^+$ state of $^{87}Rb_2$ *Phys. Rev.* A **82** 052514

[32] Brown J M and Merer A J 1979 Lambda-type doubling parameters for molecules in $\Pi$ electronic states of triplet and higher multiplicity *J. Mol. Spectrosc.* **74** 488

[33] Freed K F 1966 Theory of the hyperfine structure of molecules: Application to $^3\Pi$ states of diatomic molecules intermediate between Hund's cases (a) and (b) *J. Chem. Phys.* **45** 4214

[34] Steck D A 2010 Rubidium 87 D Line Data, revision 2.1.4, available at http://steck.us/alkalidata.

[35] Li L, Zhu Q, and Field R W 1989 Hyperfine structure of the $Na_2$ $1^3\Delta_g$ state *J. Mol. Spectrosc.* **134** 50

[36] Katô H, Otani M, and Baba M 1989 Hyperfine structure of the $Na_2$ $b^3\Pi_u$ state *J. Chem. Phys.* **91** 5124

[37] Arimondo E, Inguscio M, and Violino P 1977 Experimental determinations of the hyperfine structure in the alkali atoms *Rev. Mod. Phys.* **49** 31

[38] Bellos M A, Rahmlow D, Carollo R, Banerjee J, Dulieu O, Gerdes A, Eyler E E, Gould P L, and Stwalley W C 2011 Formation of ultracold $Rb_2$ molecules in the $v'' = 0$ level of the $a^3\Sigma_u^+$ state via blue-detuned photoassociation of the $1^3\Pi_g$ state *Phys. Chem. Chem. Phys.* **13** 18880

[39] Tsai C-C, Bergeman T, Tiesinga E, Julienne P S, and Heinzen D J 2013 Hyperfine and vibrational structure of weakly bound levels of the lowest $1_g$ state of molecular $^{87}Rb_2$ *Phys. Rev.* A **88** 052509

[40] Drozdova A N, Han X, Dai X, Wannous G, Crozet P, and Ross A J 2014 On the $2^1\Pi_g$ state of the rubidium dimer *J. Mol. Spectrosc.* **299** 25

[41] Tomza M, Goerz M H, Musiał M, Moszynski R, and Koch C P 2012 Optimized production of ultracold ground-state molecules: Stabilization employing potentials with ion-pair character and strong spin-orbit coupling *Phys. Rev.* A **86** 043424